\documentclass[preprint,amsmath,amsfonts,amssymb]{revtex4-2}

\usepackage[utf8]{inputenc}
\usepackage[T1]{fontenc}

\usepackage{xcolor}

\date{\today}

\usepackage{amsmath}
\usepackage{graphicx}
\usepackage{hyperref}
\usepackage{bm}
\newcommand{\be}{\begin{equation}}
\newcommand{\ee}{\end{equation}}
\newcommand{\ba}{\begin{eqnarray}}
\newcommand{\ea}{\end{eqnarray}}
\newcommand{\baa}{\begin{eqnarray*}}
\newcommand{\eaa}{\end{eqnarray*}}
\newcommand{\bb}{\begin{thebibliography}}
\newcommand{\eb}{\end{thebibliography}}

\newcommand{\bi}[1]{\bibitem{#1}}
\newcommand{\lab}[1]{\label{#1}}
\newcommand{\re}[1]{(\ref{#1})}

\newcommand\ve{\varepsilon}

\newcounter{my}
\newcommand{\he}%
   {\stepcounter{equation}\setcounter{my}%
   {\value{equation}}\setcounter{equation}0%
   }%
\newcommand{\she}%
   {\setcounter{equation}{\value{my}}%
    }%

\begin{document}

%\preprint{APS/123-QED}

\title{Spectral surgery and high-fidelity quantum state transfer in $XX$ chains}

\author{Luc Vinet}
 \affiliation{Centre de Recherches Math\'ematiques and Département de physique
Universit\'e de Montr\'eal, P.O. Box 6128, Centre-ville Station,
Montr\'eal (Qu\'ebec), H3C 3J7}
\author{Alexei Zhedanov}%
\affiliation{Euler International Mathematical Institute, Saint Petersburg, Russia}

%\date{\today}

\begin{abstract}
We consider an inhomogeneous $XX$ spin chain which interpolates between the Krawtchouk one with perfect state transfer and the homogeneous $XX$ chain. This model can be used to perform qubit state transfer with sufficiently high fidelity. The advantage of this model with respect to the Krawtchouk chain is that while maintaining high transfer fidelity, the coupling strengths are capped and do not become excessively large as the number of sites grows. The construction is fully analytic and is based on spectral surgery transformations of the homogeneous chain.  

\end{abstract}

%\pacs{}

%\keywords{perfect state transfer, spin chains, orthogonal polynomials}

\maketitle

\section{Introduction}
\setcounter{equation}{0}

Perfect state transfer (PST) is a protocol that performs with probability one, the transport of a qubit in an unknown state from one location to another. The design, in terms of spin chains, of devices enacting PST without error inducing external controls has been initiated some 20 years ago \cite{Bose} and is still the subject of much attention. One reason for that in the context of noisy intermediate scale quantum devices is that such quantum wires could be used instead of swap gates in circuit routing to adapt for instance to the constrained gate architectures currently available (see \cite{KVPDFC} for a Reinforcement Learning approach to such issues). 

We report here how to construct analytically a spin chain that achieves high-fidelity state transfer while avoiding the excessively large coupling strengths that arise in long chains exhibiting PST.

The basic system that is used to realize such quantum wires is still that of an $XX$ spin chain
with nearest-neighbor interactions.  The corresponding Hamiltonian $H$ is given by
 \be H=\frac{1}{2} \:
\sum_{l=0}^{N-1} J_{l+1}(\sigma_l^x \sigma_{l+1}^x + \sigma_l^y
\sigma_{l+1}^y) +  \frac{1}{2} \: \sum_{l=0}^N B_l(\sigma_l^z +1),
\lab{H_def} \ee where $J_l$ are the constants coupling the sites
$l-1$ and $l$ and $B_l$ are the strengths of  the magnetic field
at the sites $l$ ($l=0,1,\dots,N$). The symbols $\sigma_l^x, \:
\sigma_l^y,\: \sigma_l^z$ stand for the Pauli matrices which act
on the $l$-th spin.

Each spin at site $l$ has two basic states $|0 \rangle_l$ (spin down) and $|1 \rangle_l$ (spin up) such that 
$$
\sigma^z_l |0 \rangle_l = - |0 \rangle_l, \; \sigma^z_l |1 \rangle_l = |1 \rangle_l.
$$
Hence the vector space of all the chain states is spanned by the vectors
$$
|n_0, n_1, \dots, n_N \rangle = |n_0 \rangle_0 |n_1 \rangle_1 \dots |n_N \rangle_N,
$$
where each $n_l$ can take the values 0 or 1.

It is easily seen that
$$
[H, \frac{1}{2} \: \sum_{l=0}^N (\sigma_l^z +1)]=0,
$$
which implies that the eigenstates of $H$ split in subspaces
labeled by  the number of spins over the chain that are up.
In order to characterize the chains with PST, it suffices to restrict $H$
to the subspace spanned by the states which contain only one
excitation. A natural basis for that subspace is
given by the vectors
$$
|e_n \rangle = |0,0,\dots, 1, \dots, 0 \rangle, \quad n=0,1,2,\dots,N,
$$
where the only ``1" occupies the $n$-th position.  The restriction $J$ of $H$ to the 1-excitation subspace acts as follows
\be J |e_n \rangle  = J_{n+1} |e_{n+1} \rangle + B_n |e_n \rangle + J_{n}
|e_{n-1} \rangle. \lab{Je} \ee Note that \be
J_0=J_{N+1}=0 \lab{J0} \ee is assumed.

The goal is to use the chain dynamics to relocate after a time T the quantum state $|\psi\rangle=\alpha |0\rangle + \beta |1\rangle$ from the site $n=0$ to the site $n=N$. A simple analysis shows that this requires the state $|e_0 \rangle=|1,0,0,\dots, 0 \rangle$ to be unitarily evolved into the state $|e_N \rangle= |0,0,\dots, 1\rangle$, i.e. to have
\be
U(T) |e_0 \rangle = e^{i \varphi} |e_N \rangle,
\lab{UAB} \ee
where $U(t)$ is the evolution operator
\be
U(t) = \exp\left( -it H\right)
\lab{U(t)} \ee
and $\varphi$ is a real phase parameter. Condition \re{UAB} defines 
PST in $XX$ spin chains.

The initial model proposed in \cite{Bose} was a uniform $XX$ chain where all the coupling constants are the same $J_l=J, \: l=1,2,\dots,N$ and the magnetic fields are absent $B_l=0$. This model however only yields PST for chains that contain only 3 or 4 spins. For longer chains there are no times $T$ for which relation \re{UAB} is verified. It was subsequently shown \cite{Albanese},\cite{Christ} that inhomogeneous $XX$ spin chains with PST for any number of sites $N$ could be engineered by judiciously picking the coupling constants and the Zeeman terms in a non-uniform fashion. The most celebrated example is that of the Krawtchouk chain with coupling constants
\be
J_l= K \sqrt{l(N+1-l)}, \; B_l=0, \; l=0,1,2,\dots, N
\lab{J_k_K} \ee 
where $K$ is an arbitrary nonzero constant.

This model provides PST at the time $T=\pi/K$. The transfer time $T$ is independent of the chain length $N$. This does not violate the Lieb-Robinson bound \cite{LR} nor special relativity for that matter because as $N$ grows, the Hamiltonian is not uniformly bounded as the chain length increases in view of the parabolic profile of the coupling constants given by \re{J_k_K}. One can of course rescale the couplings by $N$, $J_n \rightarrow \frac{J_n}{N}$, to alleviate this issue with the result that the PST time will be $T=\frac{\pi N}{K}$. 
This fits with the bound of the speed of PST found by Yung \cite{Yung} (see also \cite{XKTN}). It is appropriate to mention in this connection the protocol developed by Xie, Tamon and Kay \cite{XKT} that allows in principle  to break that bound in the speed of PST.

Renormalizing the coupling constants by $N$ leaves however the problem that the couplings at the extremities of the chain will become very small as $N$ grows meaning that the sites towards the end of the chain are basically uncoupled. In a photonic realization of the chain with waveguides \cite{Perez} this requires those waveguides to be very far apart, an impractical situation.

In summary, the problem with the Krawtchouk chain that we are stressing lies in the fact that the ratio of the maximal value of squares of $J_l $(in the center of the chain) and the minimal one (i.e. for $l=1$) is
\be
\frac{J_{max}^2}{J_1^2} = \frac{(N+1)^2}{4N} 
\lab{rati_J} \ee
This ratio increases linearly with large $N$ which makes any useful implementation of such a chain difficult or even impossible if $N$ attains large values.

Several approaches have been proposed to achieve high-fidelity, although not perfect, state transfer while avoiding the introduction of couplings that become very large when the size of the chain grows have been explored through variations involving the uniform chain, for instance by using a number of such chains \cite{Burgarth}, or by modulating only the parameters affecting the end sites of the chain \cite{Wojcik}, \cite{Banchi}. We here consider the problem from a somewhat different angle by not insisting on the idea that the uniform chain plays a central role but by asking rather the question: is it possible to design  an analytic  chain which ``interpolates" between the Krawtchouk and homogeneous $XX$ chain, so as to be free of the above difficulties and to perform the required transfer with high fidelity?

The desired chain should satisfy the following conditions:

(i) the transfer from $|e_0 \rangle$ to $|e_N\rangle$ is ``sufficiently good",

%\vspace{4mm}

(ii) the ratio $\frac{J_{max}^2}{J_{min}^2}$ is smaller than that for the Krawtchouk chain in order to allow for a physically realistic implementation.  

We take the point of view that it is not the modulation of the parameters of the chain that is problematic, but rather the fact that these specifications become ``unbounded" in the models with PST when the size of the chain grows. Indeed, as the realization with photonic waveguides shows \cite{Perez},  it might not be much more difficult to engineer couplings with specific values at the various sites than making them all exactly equal. In that respect, we suggest that having analytic models is quite useful as this entails exact formulas for the couplings; this is a significant feature of the approach based on spectral surgery initially mentioned in \cite{VZ_PST} that we shall use in the following.

We shall thus indicate how to construct analytically these interpolating chains that have explicit expressions for couplings $J_l$ and satisfy the conditions (i) and (ii) stated above.  A noteworthy aspect of the construction is that the spectral modification is performed analytically. The resulting spin chains are not obtained through numerical optimization but through explicit spectral surgery transformations. As a consequence, closed formulas are available for the spectrum, the coupling constants and the orthogonality weights. This provides a systematic and exactly solvable framework for the design of quantum state-transfer devices.

The issue of obtaining sufficiently good transfer was also addressed in \cite{Godsil1}, where the authors introduced the notion of ``pretty good transfer" (PGT) and analyzed it in the context of the uniform chain. However, even with this weaker condition on the fidelity of the transfer, the homogeneous chain still proved to have two significant drawbacks:  (i) the set of numbers $N$ for which PGT occurs is rather limited and (ii) the time $T$ for PGT cannot be found by an efficient algorithm.

The same idea of approximately perfect transfer dubbed in this case ``almost" perfect state transfer (APST) was applied in \cite{almost} to non-uniform chains. In this case, for some models the transfer times can be explicitly computed when APST or PGT happens. These times prove finite but the corresponding models are again affected by the difficulties already pointed out for the Krawtchouk chain concerning the large values of the couplings.

This motivates the introduction of a less restrictive notion of efficient transfer. Rather than requiring perfect transfer or asymptotic notions such as APST or PGT, we shall simply look for chains whose transfer amplitude satisfies
\[
1-|A(T)|\le \delta ,
\]
where $\delta$ is a prescribed tolerance determined by the intended application. We shall refer to such transfer as {\it good-enough state transfer}. The purpose of this notion is practical rather than asymptotic: the objective is to identify analytically solvable chains that achieve high transfer fidelity while keeping the coupling strengths within realistic bounds.

With the adoption of this definition, our goals will be reached by using the spectral surgery method rooted in the Darboux transformations of orthogonal polynomials \cite{Chi} to eliminate the most non-linear part of the spectrum of the uniform chain and thereby determine analytically the desired chain.

It should be mentioned that, although in a different spirit, the approach to be followed here is not without similarity to the one offered in \cite{Kay_Inc} (see also \cite{Chen}) where it is proposed to use encoding in the outer parts of a chain to alleviate the presence of spectral points that prevent the standard PST. Taking as middle part the segment of the uniform chain that has a quasi-linear spectrum bears a resemblance in some sense to the spectral surgery that we shall perform on that chain.

The objective of the present work is not to preserve perfect state transfer at all costs, but rather to identify analytically solvable spin chains whose transfer fidelity remains high while their coupling strengths remain within a practically realistic range.

\section{Time evolution of a qubit in $XX$ spin chains}
\setcounter{equation}{0}
As shown in \cite{VZ_PST}, the dynamics of the inhomogeneous $XX$ chain can be described by using the orthonormal polynomials $\chi_n(x)$ arising from the recurrence relation
\be J_{n+1} \chi_{n+1}(x) + B_n \chi_n(x) +
J_{n} \chi_{n-1}(x) = x \chi_n(x) \lab{rec_chi} \ee
with
\be
\chi_{-1}=0, \; \chi_0=1 . \lab{ini_chi} \ee
It is convenient to introduce the monic orthogonal polynomials $P_n(x)$ through
\be
P_n(x) = J_1 J_2 \dots J_n \chi_n(x) = x^n + O\left(x^{n-1} \right).
\lab{P_monic} \ee
The orthogonality relations are of the form
\be
\sum_{s=0}^N w_s \chi_n(x_s) \chi_m(x_s) = \delta_{nm}, \lab{ort_chi} \ee
or, equivalently,
\be
\sum_{s=0}^N w_s P_n(x_s) P_m(x_s) = h_n \delta_{nm}, \lab{ort_P} \ee
where 
\be
h_n = J_1^2 J_2^2 \dots J_n^2
\lab{h_def} \ee
is the normalization constant. 
The grid points $x_s$ are the eigenvalues of the tridiagonal matrix $J$ with diagonal entries $B_l$ and off-diagonal entries $J_l$:
\be
J |x_s \rangle = x_s | x_s \rangle, \; s=0,1, \dots , N.
\lab{J_eig} \ee  
Note that the $x_s$ are nondegenerate provided that $J_l \ne0$ for $l=0,1,\dots, N$.
The discrete weights $w_s$ are given by
\be w_s =
\frac{h_N}{P_N(x_s) P_{N+1}'(x_s)}, \quad s=0,1,\dots, N,
\lab{w_s_PP} \ee
where $P_{N+1}(x)$ is the characteristic polynomial of the spectrum:
 \be
P_{N+1}(x) = (x-x_0)(x-x_1) \dots (x-x_N).  \lab{P_N+1} \ee
It is easy to show that the weights are positive $w_s>0$ and satisfy the normalization condition
\be
\sum_{s=0}^N w_s =1. \lab{norm_w} \ee
In what follows we shall take the eigenvalues $x_s$ in increasing order
\be
x_0<x_1<x_2 < \dots <x_N . \lab{incr_x} \ee
The eigenvectors $| x_s \rangle$ of the tridiagonal matrix $J$ have the expression \cite{VZ_PST} 
\be
| x_s \rangle = \sum_{n=0}^N \sqrt{w_s} \chi_n(x_s) |e_n \rangle \ . \lab{sew_expans} \ee

The tridiagonal matrix is called {\it persymmetric} if it is symmetric under reflection with respect to the main antidiagonal, i.e. if
\be
J_{N+1-l}=J_l, \; B_{N-l}=B_l
\lab{per_J} \ee
The necessary and sufficient conditions for PST are

\vspace{3mm}

(i) the matrix $J$ is persymmetric

\vspace{3mm}

(ii) the spectrum $x_s$ satisfy the conditions
\be
x_{s+1}-x_s = \kappa M_s, \: s=0,1,\dots, N-1
\lab{diff_x} \ee
where $M_s$ are positive odd integers and $\kappa$ is an arbitrary positive parameter.
If conditions (i)-(ii) are fulfilled then the minimal time $T$ for which \eqref{UAB} is satisfied is given by
\be
T = \frac{\pi }{\kappa d}, 
\lab{T_pi} \ee
where $d$ is GCD of the integers $M_0,M_1, \dots, M_{N-1}$.

The Krawtchouk $XX$ chain has the linear spectrum $x_s = K \left(s-N/2 \right), \: s=0,1,2,\dots, N$ which ensures that \eqref{diff_x}is satisfied with $M_s=1$ for all $s$ and $\kappa =K$ and the matrix $J$ defined by \eqref{J_k_K} furthermore verifies \eqref{per_J}. The corresponding polynomials $P_n(x)$ coincide with the Krawtchouk polynomials \cite{Albanese}.
Considering some other chain, assume that the matrix $J$ is still persymmetric according to \re{per_J} but that conditions \re{diff_x} do not hold. In this case PST is impossible.
However, we can suppose that the conditions \re{diff_x} are approximately satisfied. We can then expect a state transport with some possibly ``sufficient" fidelity.
It is convenient to introduce the amplitude
\be
A(t) = \langle e_N | U(t) | e_0 \rangle = \langle e_N | \exp\left( -i t H \right) | e_0 \rangle.
\lab{A_def} \ee 
Obviously for any time $t$ we have the inequality
$$
|A(t)| \le 1
$$
The PST condition \eqref{UAB}is equivalent to 
\be
|A(T)|=1
\lab{AT_cond} \ee
We say that the state transfer exhibits good-enough fidelity if 
\be
1-|A(T)| = \delta,
\lab{get} \ee
where $\delta$ is a small parameter depending on our requirements for experimental implementation. For example, for practical purposes one might take $\delta \le 0.05$. The smaller this parameter $\delta$ is, the higher the fidelity will be.

\section{Spectrally modified homogeneous $XX$ chain}
\setcounter{equation}{0}
The spectrum $x_s$ of the uniform $XX$ chain with $B_k=0, k=0,1,\dots, M$ and $J_k=1/2$ is
\be
x_s = -2 \cos \omega (s+1), \; s=0,1,\dots, M,
\lab{x_s_H} \ee
where 
\be
\omega =\frac{\pi}{M+2}. \lab{omega}
\ee
\begin{figure}[h]

%\centering

\includegraphics[width=\columnwidth]{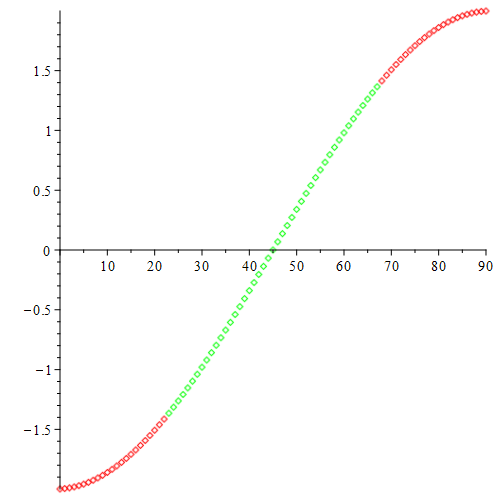}

\caption{Spectrum of the uniform $XX$ chain. The green region corresponds to the approximately linear part of the spectrum.}

%\label{fig:US}

\end{figure}
In Fig. 1 the essentially nonlinear ("red") part  of the spectrum prevents PST because conditions \re{diff_x} cannot be fulfilled.
This suggests modifying the uniform XX chain by removing the highly nonlinear "red" part of the spectrum. The remaining "green" part of the spectrum will then approximately satisfy conditions \re{diff_x} with $x_{s+1}-x_s \approx const$.  

This procedure is described in \cite{VZ_PST} and called ``spectral surgery". Spectral surgery is particularly attractive in the present context because it preserves exact solvability. Unlike generic optimization procedures for quantum state transfer, it leads to explicit expressions for the modified Jacobi matrix and hence for the coupling constants of the corresponding spin chain. More precisely the idea is the following. Assume that the spectrum of the initial homogeneous (or inhomogeneous) $XX$ chain with $M+1$ sites is $x_0< x_1<x_2< \dots < x_{M-1}<x_M$. Let $J_l$ be the corresponding parameters of this chain. In what follows we assume that magnetic fields are absent $B_l=0$.
Consider a new inhomogeneous $XX$ chain with $M-1$ sites and with spectrum $x_1<x_2< \dots < x_{M-2}<x_{M-1}$. That is, the new spectrum is obtained by removing two boundary eigenvalues $x_0$ and $x_M$. We denote the new coupling constants as  $J_l^{(1)}$ (the magnetic fields remain absent $B_l^{(1)}=0$).
Repeating this procedure step-by-step by removing the boundary eigenvalues, we arrive after $j$ iterations at the nonhomogeneous $XX$ chain with spectrum $x_j < x_{j+1} < \dots < x_{M-j-1} < x_{M-j}$ and with coupling constants $J_l^{(j)}$ and $B_l^{(j)}=0$. It is assumed that $l=0,1,2,\dots, N-1,N$, where $N=M-2j$. 

In \cite{VZ_PST} it was demonstrated that the chain with the coupling constants $J_l^{(j)}$ can be obtained from the initial chain by the successive application of $j$ Darboux transformations of the initial Jacobi matrix $J$.  In turn, these transformations are well known as Christoffel transformations and are equivalent to refactorizations of the Jacobi matrix.
In general, the coupling constants $J_l^{(j)}$ can be expressed via the initial constants $J_l$ and the values of the orthogonal polynomials $P_n(x)$ at the spectral points $x_0,x_1, \dots, x_{j-1}$ (see \cite{VZ_PST} for details). It is important to stress that the Jacobi matrix $J^{(j)}$ remains persymmetric for all $j=0,1,2,\dots$. Moreover, if the initial $XX$ chain realizes PST, then all the derived chains with Jacobi matrices $J^{(j)}$ will exhibit PST as well.   

Let $P_n^{(j)}(x), \, n=0,1,2,\dots, N$ be the set of monic orthogonal polynomials corresponding to the ``spectrally modified" Jacobi matrix $J^{(j)}$. For $j=0$ (i.e. for the case of the uniform $XX$ chain), the polynomials $P_n(x)$ coincide with the Chebyshev polynomials of second type $P_n(x)=U_n(x)$. A number of relevant things can now be said about  the associated polynomials $P_n^{(j)}(x)$.
In \cite{SZ} it was shown that the Darboux process for the Chebyshev polynomials $U_n(x)$ (which is equivalent to applying spectral surgery to the uniform $XX$ chain) leads to the so-called ``q-ultraspherical polynomials" which are well known for $q$ real\cite{KLS}. In the case of interest here, the parameter $q$ is a root of unity
\be
q=\exp\left(\frac{2 \pi i}{M+2} \right). \lab{q_M} 
\ee
The corresponding coupling constants are
\be
{J_l^{(j)}}^2= K^2 \:\frac{(1-q^l)(1-q^{l+2 j+1})}{(1-q^{l+j})(1-q^{l+j+1})}. \lab{J_q} \ee
The positive constant $K$ may be taken to be arbitrary and will depend on the concrete physical implementations. 

When $j=0$ we have the uniform chain, i.e. $J_l =K$. 
For a positive integer $j$ we have
\be
J_0=J_{N+1}=0 \lab{J_bound} 
\ee 
with $N=M-2j$. Condition \re{J_bound} means that the chain consists of $N+1$ sites: $l=0,1,\dots, N$.

From the results of \cite{SZ} it follows on the one hand, that the eigenvalues of the corresponding Jacobi matrix are (putting $K=1$ for simplicity)
\be
x_s = -2 \cos \omega (s+1+j), \quad s=0,1,\dots, N. \lab{eig_x} 
\ee 
Formula \eqref{eig_x} makes transparent the effect of the surgery
procedure: the most nonlinear part of the spectrum of the homogeneous
chain is removed while the approximately linear central portion is
retained. Since PST is associated with linear spectra, one may expect
the resulting chain to exhibit significantly improved transfer
properties while maintaining moderate coupling strengths.
On the other hand, the one-excitation spectrum of the uniform $XX$ chain with $M+1$ sites are given by \re{x_s_H}.
Hence the spectrum \re{eig_x} can be obtained from the spectrum \re{x_s_H} by removing $j$ levels from the top and $j$ levels from the bottom. In other words, this is equivalent to removing the ``red" levels in Fig. 1. This corresponds to the spectral surgery procedure described in \cite{VZ_PST}.

Recalling \eqref{omega}, one can rewrite \re{J_q} in the form
\be
J_l^2= K^2 \frac{\sin(\omega l) \sin(\omega (N+1-l))}{\cos(\omega (l-N/2)) \cos(\omega (l-N/2-1))} \lab{J_sin_cos} \ee
which involves only the integer parameters $M$ and $N$.

From Fig.2 it is seen that for a fixed number $N=100$ of spins, the profile $J_l^2$ of the coupling constants interpolates between the profile of the uniform $XX$ chain when $M=N$, (i.e. when the number of iterations is zero $j=0$) and that of the inhomogeneous Krawtchouk $XX$ chain (when $M \to \infty$). 

\begin{figure}[h]

%\centering

\includegraphics[width=\columnwidth]{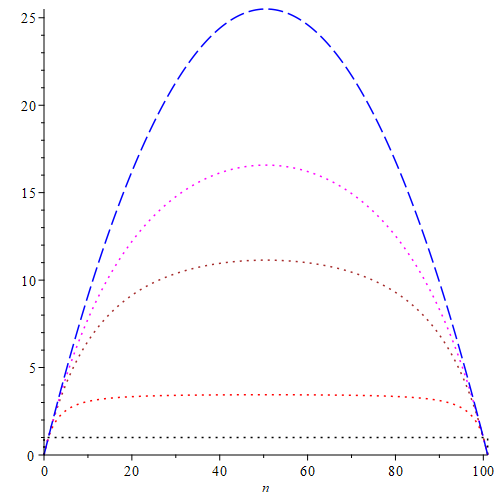}

\caption{Profiles \re{J_sin_cos} of the coupling constants of the spectrally modified $XX$ chain with $N=100$: the blue dash curve (parabola) corresponds to the Krawtchouk chain ($M \rightarrow \infty$), the black line to the uniform $XX$ chain ($j=0$) and the dots depict the spectrally modified chain with $M=110$ in red, $M=150$ in brown and $M=200$ in magenta. All plots are normalized with respect to the value $J_1$.}

%\label{fig:US}

\end{figure}

The discrete orthogonality weights are
\be
\begin{split}
& w_s (M,N)=  \\
&\kappa^{-1}  \prod_{k=0}^j \sin\left( \omega (s+j-k+1)\right) \sin\left( \omega (s+j+k+1)\right)
\end{split}
\lab{w_s_ultra} \ee
where the normalization constant is
\be
\kappa =  \frac{M+2}{2}\prod_{k=0}^j \frac{\cos\left( \omega k \right) \sin\left( \omega (2k +1)\right) }{2 \sin\left( \omega(k+1) \right)}.
\lab{kappa_MN} \ee
These weights are normalized
\be
\sum_{s=0}^N w_s =1.
\lab{norm_ws} \ee
The nonnegative integer parameter $j$ is defined as
\be
j= (M-N)/2 =0,1,2, \dots
\lab{j_MN} \ee
In particular, for $N=M$ (i.e. for $j=0$) we have the case of homogeneous $XX$ chain. Then
\be
w_s(M;M) = \frac{2}{M+2} \sin^2 \omega(s+1).
\lab{w_free} \ee
Formulas \re{w_s_ultra} and \re{kappa_MN} follow from results of \cite{SZ}. 

The ratio $R_K$ of the maximal and minimal values of $J_l^2$ for the Krawtchouk chain is
\be
R_K= \frac{J_{(N+1)/2}^2}{J_1^2} = \frac{(N+1)^2}{4N}. \lab{Kr_R} \ee     
For the corresponding ratio $R_S$ of the spectrally modified $XX$ homogeneous chain we have 
\be
R_S = \frac{J_{(N+1)/2}^2}{J_1^2} = \frac{\sin^2(\omega(N+1)/2) \cos^2(\omega(N-2)/2)}{2 \cos(\omega/2) \sin(\omega) \sin(\omega N/2)}. \lab{R_H} 
\ee
Fixing $N$ and increasing $M$ one can obtain a ratio $R_S$ that approaches $1$ (i.e. the ratio of the uniform $XX$ chain) which is more suitable as explained before. There remains to determine if the fidelity of the qubit transfer is sufficiently high.

\section{Fidelity estimation}
\setcounter{equation}{0}
Because the tridiagonal matrix $J$ is persymmetric, the amplitude $A(t)$ (recall \eqref{A_def}) of the quantum signal at the end of the chain can be calculated with the help of the following formula \cite{VZ_PST}:
\be
A(t) = \sum_{s=0}^N w_s (-1)^{N+s} \exp\left(-i x_s t \right).
\lab{A0N} \ee
Note that for $t=0$ we have
\be
A(0)= \sum_{s=0}^N w_s (-1)^{N+s}=0,
\lab{A0N=0} \ee
a consequence of the properties of persymmetric matrices \cite{persym}. Formula \re{A0N=0} means that at $t=0$, the quantum signal (qubit) is concentrated at $n=0$ and that hence the amplitude at $n=N$ is zero.
Given the explicit expression of the weights \re{w_s_ultra}, one can evaluate the fidelity defined in \eqref{get}
\be
\delta(T) = 1-|A(T)|
\lab{F} \ee
for different values of $N,M,T$. The quantity $\delta(T)$ should not be viewed as defining a new asymptotic notion analogous to APST or PGT. It is simply a practical measure of the transfer error for a finite chain and a finite transfer time. Our goal is to identify analytically solvable chains for which $\delta(T)$ remains sufficiently small while the coupling strengths stay within a realistic range. The main problem is:  find the time $T$ such that $\delta(T)$ has the smallest (necessarily positive) value for the given parameters $N$ and $M$. Remember that smaller and smaller $\delta (T)$ will amount to higher and higher fidelities that can be defined by $1-\delta(T)$.

We present several estimates for $N=100$.
It is convenient to normalize the spectrum of the uniform $XX$ chain (and hence its Hamiltonian) according to
\be
x_s = -2(M+2) \cos \omega (s+1), \: s=0,1,\dots , M
                                        \lab{x_s_n} \ee   
with $\omega = \pi/(M+2)$. Then for sufficiently large values of $M$ the levels $x_s$ in the middle part of the spectrum 
\re{x_s_n} have a linear behavior with $x_{s+1}-x_s \approx 2 \pi$. We know that the PST time of the Krawtchouk chain (corresponding to $M \to \infty$) is $T_K = 1/2$. 
We can then use this $T_K$ as a ``zeroth order approximation" for the time $T$ which yields a minimal value of $\delta(T)$, i.e. we will search for a $T$ of the form
\be
T = 1/2 + \ve.
\lab{T=1/2+} \ee
Numerical evaluation of formula \re{A0N} gives the following results

\vspace{2mm}

(i) for $N=100$ and $M=110$ we have $\ve = 10^{-2}$ and $\delta(T) \approx 0.13$. Such a fidelity is an improvement over that of the uniform $XX$ chain, although it may still be insufficient for some experimental implementations.

\vspace{2mm}

(ii) for $N=100$ and $M=120$ we have $\ve = 10^{-2}$ and $\delta(T) \approx 0.05$. This level of fidelity may already be adequate for practical implementations. Moreover, in this case the ratio $R_S$ between the maximal and minimal values of $J_l^2$ is approximately 5, while for the Krawtchouk chain this ratio is 25. This means that this chain is much better behaved.

\vspace{2mm}

(iii) for $N=100$ and $M=150$ we have $\ve = 0.005$ and  $\delta (T) \approx 0.008$. This fidelity is sufficiently close to unity for most practical purposes.

\vspace{2mm}

Finally notice that increasing $N$ with a fixed ratio $M/N$ one can achieve very good fidelity. Consider, for instance the values $M=550, N=500$. In this case $\delta (T) \approx 0.07$. For $M=1100$ and $N=1000$, $\delta(T) \approx 0.05$ which is already ``good enough" and the ratio $R_S=25$ is 10 times smaller than the ratio $R_K=250$ for the Krawtchouk chain that generates PST. This means that for long spin chains the spectrally modified chain offers practical candidates as possible registers for quantum computers or as tools to help with circuit routing.

\section{Conclusion}

We have shown that spectral surgery of the homogeneous $XX$ chain generates a family of analytically solvable spin chains that interpolate between the uniform chain and the Krawtchouk chain. The construction preserves explicit formulas for the spectrum, coupling constants and orthogonality weights, thereby providing a rare example of a nontrivial exactly solvable family of quantum communication channels obtained through controlled spectral modification.

The construction preserves explicit control over the spectrum and coupling strengths while significantly reducing the variation between the largest and smallest couplings when compared with the Krawtchouk model. Although perfect state transfer is generally lost, numerical estimates show that high transfer fidelities can nevertheless be achieved.

These spectrally modified chains therefore provide analytically tractable and physically realistic candidates for quantum communication channels and quantum-routing architectures in situations where bounded coupling strengths are more important than exact perfect state transfer.

\bigskip\bigskip
{\Large\bf Acknowledgments}
\bigskip

AZ thanks the Centre de Recherches Math\'ematiques (Universit\'e de
Montr\'eal) for hospitality.  The authors would like to thank
M. Christandl, M.Derevyagin and A.Filippov for stimulating
discussions. 
The research of LV is supported in part by a research grant from the Natural Sciences and Engineering Research Council
(NSERC) of Canada.

\newpage

\bb{99}

\bi{Albanese} C. Albanese, M. Christandl, N. Datta, A. Ekert, {\it
Mirror inversion of quantum states in linear registers},  Physical
Review Letters {\bf 93} (2004), 230502.

\bi{Banchi} L. Banchi, T. J. G. Apollaro, A. Cuccoli, R. Vaia, and P. Verrucchi, {\it Optimal dynamics for quantum-state and entanglement transfer through homogeneous quantum systems}, Physical Review A {\bf 82}, 052321 (2010). arXiv:1006.1217v1.

\bi{Bose} S.~Bose, {\it Quantum communication through spin chain
dynamics: an introductory overview},  Contemporary Physics, {\bf 48},
(2007), 13 -- 30.

%\bi{Bruderer} M. Bruderer, K. Franke, S. Ragg, W. Belzig, D. Obreschkow, {\it Exploiting boundary states of imperfect spin chains %for  high-fidelity state transfer}, Phys. Rev. A {\bf 85} (2012), 022312.  arXiv:1112.4503

\bi{Burgarth} D. Burgarth, V. Giovannetti and S. Bose, {\it Efficient and perfect state transfer in quantum chains}, Journal of Physics A  Mathematical and Theoretical,{\bf 38} , 6793 (2005). arXiv:quant-ph/0410175v3

%\bi{BB} D.Burgarth S.Bose, {\it Conclusive and arbitrarily perfect quantum state transfer using parallel spin chain channels}, Phys. %Rev. A
%{\bf 71}, (2005) 052315 . arXiv: quantum-ph/0406112v4.

%\bi{bounds} C.Burrell and T. Osborne, {\it Bounds on information propagation in disordered quantum spin chains}, Phys. Rev. Lett. %{\bf 99}, (2007), 167201  ArXiv: quant-ph/0703209v3

%\bi{disorder} C.Burrell, J.Eisert and T.Osborne, {\it Information propagation through quantum chains with fluctuating disorder}, %Phys. Rev. A {\bf 80} (2009), 052319 . ArXiv: 0809.4833v1.

%\bi{fractal} G. De Chiara, D. Rossini, S.Montenegro and R. Fazio, {\it From perfect to fractal transmission in spin chains}, Phys. %Rev. A {\bf 72}, 012323 (2005) arXiv: quant-ph/0502148v2.

\bi{Chen} X. Chen, R. Mereau, and D. L. Feder, {\it Asymptotically perfect efficient quantum state transfer across uniform chains with two impurities}, Physical Review A {\bf 93}, 012343 (2016). arXiv:1511.00038v1.

\bi{Chi} T. Chihara, {\it An Introduction to Orthogonal
Polynomials}, Gordon and Breach, NY, 1978.

\bi{Christ} M. Christandl, N. Datta, T. C. Dorlas, A. Ekert, A. Kay and A. J. Landahl, {\it Perfect transfer of arbitrary states in quantum spin networks}, Physical Review A {\bf 71} (2005), 032312. arXiv:quant-ph/0411020

\bi{persym} V. Genest, S. Tsujimoto, L. Vinet and A. Zhedanov, {\it Persymmetric Jacobi matrices, isospectral deformations and orthogonal polynomials}, Journal of Mathematical Analysis and Applications , {\bf 450} (2017), 915--928. arXiv:1605.00708.

%\bi{Godsil} C.Godsil, {\it State Transfer on Graphs}, Discrete Math. {\bf 312}(1): 129--147 (2012). arXiv:1102.4898

\bi{Godsil1} C. Godsil, S. Kirkland, S. Severini, J. Smith, {\it Number-theoretic nature of communication in quantum spin chains}, Physical Review Letters {\bf 109} (2012), 050502; arXiv:1201.4822.

%\bi{GVZ} A.Gr\"unbaum, L.Vinet and A.Zhedanov, {\it Birth and
%death processes and quantum spin chains}, arXiv:1205.4689.

\bi{Kay_Inc} A. Kay, {\it Incorporating Encoding into Quantum System Design}, Physical  Review A {\bf  109}, 042408. arXiv:2207.01954v2.

%\bi{Kay1} A.Kay, {\it Perfect State Transfer: Beyond Nearest-Neighbor Couplings}, Phys. Rev. A {\bf 73} (2006), 032306 . ArXiv: %quant-ph/0509065v2.

%\bi{Kay} A.Kay, {\it A Review of Perfect State Transfer and its
%Application as a Constructive Tool}, Int. J. Quantum Inf. {\bf 8}
%(2010), 641--676;  arXiv:0903.4274.

%\bi{Karbach} P.Karbach and J.Stolze, {\it Spin chains as perfect
%quantum state mirrors},  Phys.Rev.A {\bf 72} (2005), 030301(R)

\bi{KLS} R. Koekoek, P. Lesky, R. Swarttouw, {\it Hypergeometric
Orthogonal Polynomials and Their Q-analogues},  Springer-Verlag,
2010.

\bi{KVPDFC} D. Kremer, V. Villar, H. Paik, I. Duran, I. Faro, J. Cruz-Benito, {\it Practical and efficient quantum circuit synthesis and transpiling with Reinforcement Learning}, arXiv:2405.13196, 2024. 
%\bi{Lang}. S.Lang, {\it Introduction to Diophantine Approximations}, Springer-Verlag, 1991.

%\bi{LZ} B.M.Levitan and V.V.Zhikov, {\it Almost periodic functions and differential equations}, Cambridge University Press, 1982.

\bi{LR} E. H. Lieb and D. W. Robinson , {\it The Finite Group Velocity of Quantum Spin Systems}, Communications in
Mathematical Physics {\bf 28} (1972), 251--257.

%\bi{MKE} C.Marletto, A.Kay and A.Ekert, {\it How to counteract systematic errors in quantum state transfer}, arXiv: 1202.2978v1.

%\bi{Shi} T. Shi, Y.Li , A.Song, C.P.Sun,  {\it Quantum-state
%transfer via the ferromagnetic chain in a spatially modulated
%field}, Phys. Rev. {\bf A 71} (2005), 032309, 5 pages,
%quant-ph/0408152.

\bi{Perez} A. Perez-Leija, R. Keil, A. Kay, H. Moya-Cessa, S. Nolte, L.-C. Kwek, B. M. Rodriguez-Lara, A. Szameit, D. N. Christodoulides, {\it Coherent quantum transport in photonic lattices}, Physical Review A {\bf 87}, 012309 (2013)

%\bi{RSA} R.Ronke, T.Spiller, I.D'Amico, {Long-range interactions and information transfer in spin chains}, J. Phys.: Conf. Ser. {\bf %286} (2011), 012020.  arXiv: 1101.4509v1

\bi{SZ} V. Spiridonov and A. Zhedanov, {\it q-Ultraspherical polynomials for q a root of unity}, Letters in Mathematical Physics
{\bf 37} (1996), 173--180. arXiv:q-alg/9605033v1

%\bibitem{Sz} G. Szeg\H{o}, Orthogonal Polynomials, fourth edition,  AMS, 1975.

%\bi{Wang} Y.~Wang, F.~Shuang, H.~Rabitz,  {\it All possible
%coupling schemes in XY spin chains for perfect state transfer},
%Phys. Rev. {\bf A 84}, (2011) 012307, arXiv:1101.1156.

\bi{VZ_PST} L. Vinet, A. Zhedanov, {\it How to construct spin chains with perfect spin transfer}, Physical Review A {\bf 85} (2012), 012323.

\bi{almost} L. Vinet, A. Zhedanov, {\it Almost perfect state transfer in quantum spin chains}, Physical Review A {\bf 86}, 052319 (2012)

\bi{Wojcik} A. W\'ojcik, T. Luczak, P. Kurzynski, A. Grudka, T. Gdala, and M. Bednarska, {\it Unmodulated spin chains as universal quantum wires}, Physical  Review {\bf A 72}, 034303 (2005). arXiv:quant-ph/0505097v1.

\bi{XKTN} W. Xie, A. Kay, and C. Tamon, {\it A Note on the Speed of Perfect State Transfer}, arXiv:1609.01854.

\bi{XKT} W. Xie, A. Kay, and C. Tamon, {\it Breaking the Speed Limit for Perfect Quantum State Transfer}, Physical Review A {\bf  108}, 012408 (2023)

\bi{Yung} M.-H. Yung, {\it Quantum speed limit for
perfect state transfer in one dimension}, Physical Review A {\bf  74}, 030303(R) (2006).

\eb

\end{document}